\documentclass[12pt,floatfix,pre]{revtex4}
\usepackage{psfrag}
\usepackage{graphicx}
\usepackage{dcolumn}
\usepackage{bm}
\usepackage{natbib}
\usepackage{amsmath} 
\begin{document}

% marginal note macros of JE - use \nw{new material}
\def\strutdepth{\dp\strutbox}
\def\nw#1{\strut\vadjust{\kern-\strutdepth\vtop to0pt{\vss\hbox to\hsize
{\hskip\hsize\hskip5pt$\leftarrow$\hss\strut}}}{\em #1}}

\title{Instability of a moving contact line}
\author{Jens Eggers}

\affiliation{
School of Mathematics, 
University of Bristol, University Walk, \\
Bristol BS8 1TW, United Kingdom 
        }
\begin{abstract}
We study a solid plate plunging into or being withdrawn 
from a liquid bath, to highlight the fundamental difference between
the local behavior of an advancing or a receding contact line, 
respectively. It is assumed that the liquid partially wets the solid,
making a finite contact angle in equilibrium. In our hydrodynamic 
description which neglects the presence of the outer gas
atmosphere, an advancing dynamic wetting line persists to 
arbitrarily high speeds. The receding wetting line, on the 
other hand, vanishes at a critical speed set by the competition
between viscous and surface tension forces. In the advancing case,
we apply existing matching techniques to the plunging plate geometry,
to significantly improve on existing theories.
For the receding contact line, we demonstrate for the first time
how the local contact line solution can be matched to the far-field meniscus.
In doing so, we confirm our very recent criterion for the instability
of the receding contact line. The results of both the advancing and the 
receding cases are tested against simulations of the full model equations.
\end{abstract}

\maketitle
\section{Introduction}
A number of recent experiments \cite{Q91,SP91,PFL01,SK03} have
tested the stability of forced advancing and receding contact lines
under conditions of partial wetting.
For example, if a solid plate or fiber is plunging into
a liquid bath to be coated (advancing contact line), the speed can be 
quite high (m/s) \cite{SK03}, while maintaining a stationary contact line.
In the opposite case of withdrawal (receding contact line) \cite{SP91}, 
a stationary contact line is observed only for very low speeds, 
and a macroscopic film is deposited \cite{SP91,Q99} typically at a speed of 
only a few cm/s. 

The description of a moving contact line is complicated by the fact that
the Navier-Stokes equation with standard no-slip boundary 
conditions \cite{LL84} breaks down near it, because such a hypothetical
flow would produce an infinite energy dissipation \cite{HS71}.
Instead, some microscopic length scale $\lambda$ must be invoked that cuts 
off this singularity, which in this paper we are going to take as a 
slip length. As a result, the local flow near the contact line is 
characterized by a typical length of about a nanometer \cite{TR89},
which has to be matched \cite{H} to the macroscopic flow away from the 
solid. In this paper, we apply a matching method developed for spreading 
drops \cite{H83} to the plunging plate, and test the result by comparing
to numerical simulations. This very significantly improves the results
of earlier calculations for the same problem \cite{GHL90,H93}.
For the opposite case of a receding contact 
angle we find that a new matching procedure is needed. The results confirm
our very recent criterion \cite{E042} for the instability of a receding
contact line. 

\begin{figure}
\begin{minipage}[t]{8cm}
 \includegraphics[width=1.\textwidth]{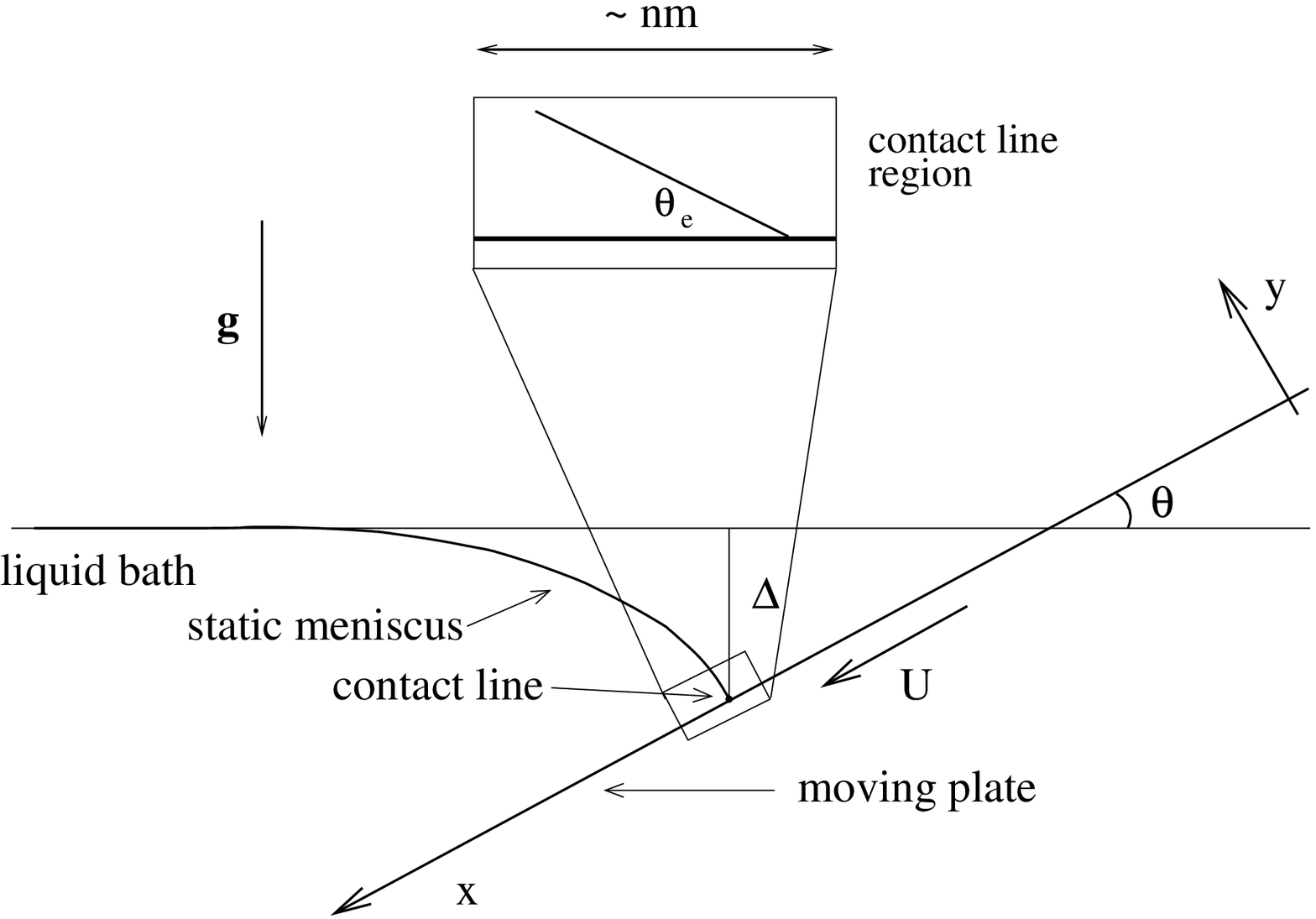}
\end{minipage}
\hfill
\begin{minipage}[t]{8cm}
    \includegraphics[width=1.\textwidth]{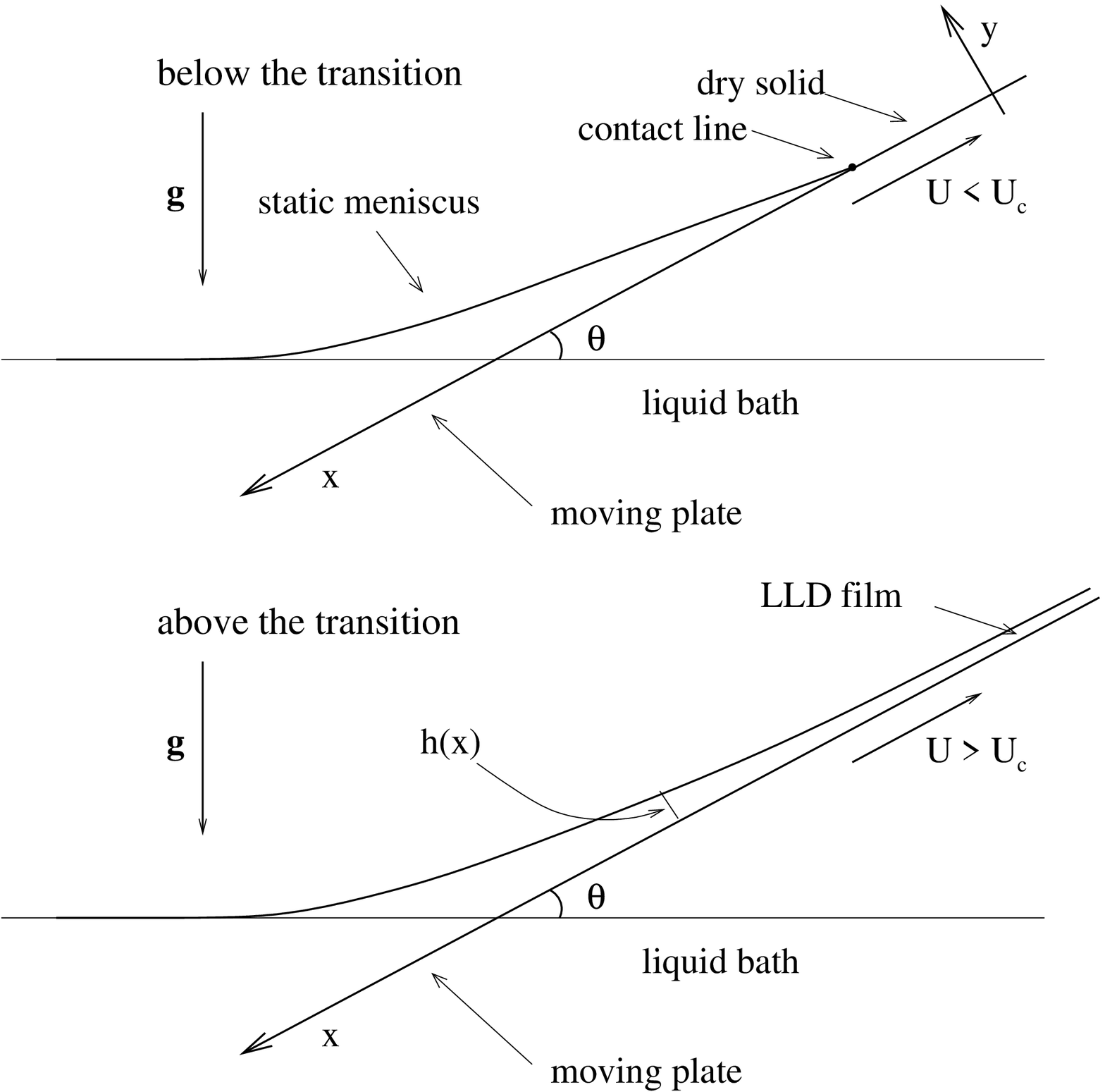}
\end{minipage}
\hfill
\caption{\label{setup} 
A schematic of the setup: a plate is being withdrawn at an angle $\theta$
on the right, and being pushed into the fluid on the left. 
Only the flow to the left of the plate is considered, which is assumed
long enough for the coupling to the other side to be ignored. 
At the contact line the microscopic slope of the 
interface is $h'(0) = \theta_e$. 
Since the interface is highly curved, this microscopic contact angle 
is only observed on a scale of nanometers (expanded region on the 
left). The depression of the advancing contact line (left) relative to the 
level of the liquid bath is $\Delta$. The contact line is stable at any speed, 
while for the receding contact line (right) there exists a 
critical speed $U_c$ above which the contact line vanishes. Instead, 
the plate is covered by a thin film. 
   }
\end{figure} 

Figure \ref{setup} illustrates the geometry to be considered in the present
paper. On the left, a solid plate is pushed into a pool of viscous 
liquid. As a result, the interface deforms and the contact line 
is pushed downward relative to its equilibrium position. Within the 
present model, this advancing contact line is stable at {\it any} speed.
If the plate is withdrawn from the bath, fluid is 
pulled up with the plate, and a new contact line position is 
established. However, this state is realizable only at speeds 
below a critical speed $U_c$. For $U>U_c$ the contact line is 
no longer sustainable and continues to move up the plate \cite{H01}.
In the stationary state, the plate is covered by a thin film, 
first described by Landau, Levich, and Derjaguin (LLD) \cite{LL42,D43}.
To understand the fundamental difference between pushing a plate and 
pulling it out, one has to consider the matching between the 
region very close to the contact line, and the capillary profile
away from it. Qualitatively, the difference in the behavior 
of the advancing and of the receding contact line makes sense: if the
plate is pushed into the liquid, the interface is bent away from the 
solid, and viscous forces are reduced. If the plate is pulled, the 
interface is pulled toward the solid, making the film thinner and 
enhancing viscous effects. Hence in the latter case there is a positive
feedback increasing viscous forcing and thus leading to instability. 

The strong energy dissipation near a moving contact line results 
from the fact that viscous forces become very large as the thickness
of the liquid film goes to zero \cite{HS71}. 
As a result of the interplay between viscous 
and surface tension forces, the interface is highly curved, and 
the contact line speed $U$ is properly measured by the capillary number
$Ca = U\eta/\gamma$, where $\eta$ is the viscosity of the fluid and 
$\gamma$ the surface tension between fluid and gas. Owing to this 
bending the interface angle measured at, say, $100 \mu$m
away from the contact line differs \cite{G85,K93,FJ91,MGD93} significantly
from the microscopic angle directly at the contact line. 

The existence of a slip length $\lambda$, which is of the order of a few 
molecular diameters under normal circumstances \cite{TT97,CLR04,LRM04},
implies a convenient separation of length scales between 
two different parts of the surface profile. On one hand, there is 
a contact line region whose typical scale is $\lambda$, on the other
hand there is an ``outer'' meniscus region, whose typical length 
scale is set by the capillary length $\ell_c=\sqrt{\gamma/(\rho g)}$.
Thus one expects the local ``inner'' behavior of the profile near the 
contact line to be of the form
\begin{equation}  
\label{scal}  
h_{in}(x) = 3\lambda H\left(\frac{x\theta_e}{3\lambda}\right), 
\quad \xi = \frac{x\theta_e}{3\lambda},
\end{equation}  
since $\lambda$ is the only available length scale. 
The dependence on the equilibrium contact angle $\theta_e>0$ 
was introduced for later convenience. 

Firstly, one finds from (\ref{scal}) that the curvature of the
interface is $h''(x)=\theta_e^2H''(\xi)/(3\lambda)$, which becomes large for 
$\lambda \rightarrow 0$ as expected, while the slope is 
$h'(x)=\theta_e H'(\xi)$. The inner solution (\ref{scal}) has to 
be matched to an outer solution $h_{out}(x)$, whose curvature 
is of the order of $\ell_c^{-1}$, which means we have to join the two
solutions at some scale $\epsilon$ with 
$\ell_c \gg \epsilon \gg \lambda$. This implies
that the argument of $h_{out}$ can effectively be taken at 
$x=0$, while $\epsilon\theta_e/(3\lambda) \gg 1$. Thus the matching
condition is 
\begin{equation}  
\label{match_cond}  
h''_{out}(0) = \theta_e^2 H''(\infty)/(3\lambda),
\end{equation}  
which ensures that the inner and the outer solutions are 
compatible. We will see that (\ref{match_cond}) needs to be supplemented by 
what is essentially a condition for the slope. 

The matching condition (\ref{match_cond}) is the key to 
understanding wetting behavior. Let us summarize the main 
results of this paper by analyzing the solution qualitatively 
for the two cases of an advancing contact angle (plate 
plunging into the fluid) and of a receding contact angle (plate 
being withdrawn). For very small $\lambda$, it is clear
that $H''(\infty)$ must be small for matching to be possible, 
so in the limit, $H''(\infty)= 0$ becomes the boundary condition for
the inner problem. We will see that for an advancing contact angle 
this boundary condition yields the inner scaling function $H(\xi)$,
first found by Voinov \cite{V76}. The matching to Voinov's solution 
is slightly complicated by the presence of {\it logarithmic} terms 
in the slope \cite{H83}. 

On the other hand, Voinov's solution cannot be 
applied to the case of a {\it receding} contact line.
Rather, all inner solutions maintain a {\it finite} curvature 
$H''(\infty) > 0$. This means that at too small a value of $\lambda$
the matching condition (\ref{match_cond}) can no 
longer be obeyed and the inner and the outer solutions are
incompatible. As a result, the contact line vanishes. In a typical 
experiment, $\lambda$ is constant and the {\it speed} is increased,
but the effect is the same: since the curvature of the interface 
is caused by viscous forces \cite{HS71}, $H''(\infty)$ increases
with speed and matching becomes impossible. This explains the
phenomenology described in Fig.\ref{setup} above. The impossibility
of matching a receding contact line above a critical capillary number
was already noticed in \cite{V00}, based on numerical integration of the 
thin-film equations. Analytical solutions for the inner solution will
permit us to give a much more complete description.

In the next section we will introduce the hydrodynamic equations
to be used for the calculation of stationary profiles. 
We will confine ourselves to the ``lubrication approximation'',
valid in the limit that the liquid film is thin. In the following
section we consider the case of a solid plate being pushed
into the liquid (advancing contact angle). By matching an inner
to an outer solution, we compute the profile as a function of speed. 
In the fourth section we introduce the matching procedure 
for the opposite case of a plate being withdrawn (receding
contact angle). The failure of this matching procedure gives 
the critical capillary number at which the contact line can no longer
exist, in agreement with our earlier result \cite{E042}. 
For speeds below the critical value we again find the interface
profile. All our analytical results are tested by comparison 
with numerical solutions of the original equations. 
Finally, we summarize our results and indicate directions of 
future research. 

\section{Lubrication description}

As illustrated in Fig.\ref{setup}, we are considering a plate being
pushed into or being withdrawn from a liquid bath at an angle $\theta$.
This means we have to solve the steady Navier-Stokes equation with a free 
surface, and no-slip boundary conditions on the plate. Since the plate
is moving with speed U, the contact line between fluid,
solid, and gas is {\it moving relative to the solid}. As explained above,
the Navier-Stokes equation does not allow such a solution, and a
small-scale cutoff has to be introduced at the contact line \cite{G85,K93}.
The dominant mechanism responsible for cutoff depends on the 
particular system under study \cite{K93,ES04}. 
As a representative example, we assume that the corner singularity 
is relieved by allowing the fluid to slip across the solid surface,
since the matching to the slip region of size $\lambda$ is well
understood in this case \cite{H83}. According to the Navier slip 
law \cite{HS71}, the fluid speed relative to the solid is 
proportional to the shear rate:
\begin{equation}  
\label{Navier}  
u(x,0)-U = \lambda\frac{\partial u}{\partial y} \quad \mbox{at} \quad y=0.
\end{equation}
Recently, we found \cite{E041} that various modifications of 
(\ref{Navier}) have a minimal influence on the interface away from the 
contact line. Hence we do not believe that the particular cutoff
mechanism used is of great importance.

A much thornier issue is the slope $h'(0)$ of the fluid
layer to be specified at the contact line. It is well 
appreciated that molecular processes are involved \cite{BH69,PRG99}, 
which are beyond a hydrodynamic description. This will lead to a 
an effective {\it speed dependence} of the contact angle as defined
on a scale of nanometers. The importance of these microscopic 
effects relative to hydrodynamic ones is determined by the amount
of energy dissipation involved in either process \cite{BG92}. 
Thus for high viscosities the speed dependence of the microscopic 
angle can most likely be ignored, in agreement with experimental data 
\cite{H75}. If there is no ``intrinsic'' speed dependence of the 
microscopic angle, it must coincide with its equilibrium value 
$\theta_e$ at {\it zero} speed. We thus take $h'(0)=\theta_e$. 

The mathematical problem simplifies significantly if we assume
that the angle the interface makes with the solid is always small, 
and that viscosity is sufficiently large for inertia to be 
ignored. In this limit, one
can find an approximate description of the hydrodynamic equations
which eliminates the flow field \cite{H83,H01}, and the 
so-called ``lubrication equation'' can be written entirely
in terms of the free-surface profile $h(x)$.
Non-dimensionalizing all lengths with the
capillary length $\ell_c=\sqrt{\gamma/(\rho g)}$ one 
finds \cite{H01}
\begin{equation}  
\label{lub}  
\frac{\pm3Ca}{h^2 + 3\lambda h} = h''' - h' + \theta,
\end{equation}  
where we consistently used the small-angle approximation
$\tan(\theta)\approx\theta$. To distinguish more clearly between
advancing and receding contact lines we always take $Ca$ as a 
{\it positive} quantity, and rather change the sign in the equation.
The - sign corresponds to the plate plunging into the 
liquid, the + sign to the opposite case of the plate being withdrawn.

Viscous forces appear on the left of (\ref{lub}) (proportional to the speed),
and diverge quadratically as $h$ goes to zero at the contact line. 
As a result, viscous dissipation would diverge if it were not for
the presence of slip, which weakens the singularity.
Near the contact line the surface is highly curved, so the
first term on the right of (\ref{lub}), which comes from
surface tension, balances the viscous term on the left. 
The other two terms stem from gravity and only come into play
at greater distances from the contact line. The film thickness vanishes 
at the contact line, where the slope is $h'(0)=\theta_e$ as discussed above.
Far away from the plate the surface coincides with the 
liquid bath, so the third boundary condition is $h'(\infty)=\theta$.
Rescaling the layer thickness $h$ with the equilibrium contact 
angle $\theta_e$, one finds that there remain {\it three} parameters 
in the problem, namely the combinations $Ca/\theta_e^3$, $\lambda/\theta_e$, 
and $\theta/\theta_e$. However, we find it more intuitive to keep the 
original parameters, and to state results in terms of the above
combinations if convenient. 

\section{Pushing}

We begin by considering a plate being pushed into a viscous fluid,
corresponding to the - sign in (\ref{lub}). 
In the spirit of the matching condition (\ref{match_cond}) 
we approach this problem by first considering the leading-order 
behavior near the contact line, where $h$ goes to zero. As discussed
above, this equation \cite{H83} is 
\begin{equation}  
\label{cl}  
\frac{-3 Ca}{h^2 + 3\lambda h} = h''', 
\end{equation}  
which we studied in detail in \cite{E041} for a more general class of 
slip models. The characteristic
scale of the local solution is the slip length $\lambda$, so it is 
convenient to introduce the scaled variables (\ref{scal}),
which leads to 
\begin{equation}  
\label{sim}  
\frac{\delta}{H^2 + H} = -H''',
\end{equation}  
where $\delta=3Ca/\theta_e^3$ is the rescaled capillary number. 

As argued above, the matching condition (\ref{match_cond}) leads
to the requirement that the curvature $H''(\xi)$ {\it vanishes} 
for large $\xi$ as the limit $\lambda\rightarrow 0$ is performed.
This means that the boundary conditions for the solution of (\ref{sim}) are
\begin{equation}  
\label{bc}  
H(0) = 0, \quad H'(0)=1, \quad H''(\infty)=0.
\end{equation}  
The only parameter now appearing in the problem is the rescaled 
capillary number $\delta$, and 
equations (\ref{sim})-(\ref{bc}) uniquely specify the profile
close to the contact line. 

This inner solution can be found by expanding in a power series in 
the capillary number in a manner described in many papers  
\cite{H83,H92,E041}. If one writes the solution in terms of 
$h'^3$, its behavior for large $x/\lambda$ can be written as
\begin{equation}  
\label{voinov}  
h_{in}'^3(x) - \theta_e^3 = 9Ca\ln(x/L)
\end{equation}  
to any order in the capillary number \cite{E041}. 
This is the form originally proposed by 
Voinov \cite{V76}, using more qualitative 
arguments. The length $L$ appearing inside the logarithm 
can be computed as a power series in the capillary number:
\begin{equation}  
\label{L}  
L = \frac{3\lambda}{e \theta_e}
\left[1 - \frac{\pi^2-1}{2\theta_e^3}Ca + O(Ca^2)\right],
\end{equation}  
but for simplicity we only take the leading order term into account 
here, as corrections introduced by the next order are usually 
quite small \cite{E041}.

The crucial point of representing the inner solution 
in the form (\ref{voinov}) is that the only parameter multiplying
$\ln(x)$ is the capillary number, which is defined in terms of the 
``outer'' problem. Namely, the outer problem (\ref{voinov}) needs to 
be matched to is 
\begin{equation}  
\label{out}  
\frac{-3 Ca}{h_{out}^2} = h_{out}''' - h_{out}' + \theta,
\end{equation}  
which does not contain any contact line parameters like 
$\lambda$ or $\theta_e$. Owing to the strong singularity
for $h\rightarrow 0$ \cite{HS71,H83}, (\ref{out}) does not have 
a solution with finite slope at the contact line $h_{out}(0)=0$,
making it impossible to impose a slope.
Instead, we are seeking a solution of (\ref{out}) that has the 
form (\ref{voinov}) for $x\rightarrow 0$, in other words
\begin{equation}  
\label{match}  
h_{out}'^3(x) = 9Ca\ln(x) + F , \quad x\rightarrow 0,
\end{equation}  
where $F$ is a constant to be computed. 
As demonstrated 
in \cite{H83}, this can be achieved by expanding the outer solution
in a power series in $Ca$:
\begin{equation}  
\label{exp}  
h_{out}(x) = h_0(x) + Ca h_1(x) + O(Ca^2). 
\end{equation}  

The equation for $h_0$, representing a balance of surface tension 
and gravity, is 
\begin{equation}  
\label{h0}  
h_0''' - h_0' + \theta = 0.
\end{equation}  
Equation (\ref{h0}) has the following family of solutions, which are
finite at infinity and which vanish at the contact line:
\begin{equation}  
\label{h0sol}  
h_0(x) = \theta x + (\theta - \theta_{ap})(e^{-x}-1).
\end{equation}  
The slope $h'_{out}(0) = \theta_{ap}$ of (\ref{h0sol}) at
the contact line is called the ``apparent'' contact angle
and is a free parameter still to be determined by the matching
procedure. Its name is motivated by the fact that a macroscopic 
measurement of the profile on the scale 
$\ell_c$ will yield a profile close to (\ref{h0sol}). 
If extrapolated to the contact line, the angle will appear to 
be $\theta_{ap}$, rather than the true microscopic value $\theta_e$.
Obviously, (\ref{h0sol}) cannot be matched directly to the contact 
line solution (\ref{voinov}), which contains a logarithm.
This logarithmic dependence will come out of the next-order solution
$h_1$, whose equation reads
\begin{equation}  
\label{h1}  
h_1''' - h_1' = f(x), \quad f(x) = -3/h_0(x)^2. 
\end{equation}  

The solution is straightforward \cite{BO78}:
\begin{equation}  
\label{h1sol}  
h_1(x) = -\frac{e^x}{2}\int_1^{\infty} e^{-t}f(t) dt
+ \int_1^x (\cosh(t-x)-1)f(t) dt + K e^{-x} + K_2, 
\end{equation}  
where once more any particular solution that is growing at
infinity was suppressed, and $K,K_2$ are constants of integration.
To find $K,K_2$, we note that $h$ must vanish at the contact line,
giving the first condition $h_1(0) = 0$. To compute $h_1(0)$, we observe
that the second integrand of (\ref{h1sol}) can be expanded like
$\cosh(t-x)-1 = \cosh(t)-1 - x\sinh(t) + O(x^2)$. The term linear 
in $x$ does not contribute for small $x$, since 
$\sinh(t)f(t)$ behaves like $1/t$ for small arguments, and hence
\[
\lim_{x\rightarrow 0} x\int_1^x \sinh(t)f(t)dt = 0.
\]
Thus one finds 
\begin{equation}  
\label{cc}  
0 = h_1(0) = -\frac{1}{2}\int_1^{\infty} e^{-t}f(t) dt
+ \int_0^1 (1-\cosh(t))f(t) dt + K + K_2,
\end{equation}  
where the singularity of $f(t)\propto1/t^2$ cancels out to make 
the second integral convergent.

Next we find a condition at infinity by noting that 
\begin{equation}  
\label{inf}  
h(x) = \theta x + \Delta + O(1/x),
\end{equation}  
where $\Delta$ is vertical distance of the contact line from the 
undisturbed surface. Since we want $h_0$ to represent the far-field
behavior of the profile, we put $h_1(\infty) = 0$, and thus 
$\Delta=\theta_{ap}-\theta$, by comparison with (\ref{h0sol}).
This means we have
\[
0 = h_1(\infty) = \lim_{A\rightarrow\infty}\left\{
\frac{e^{-A}}{2}\int_1^Ae^tf(t)dt\right\} - \int_1^{\infty}f(t)dt + K_2.
\]
The first of the two terms on the right is zero, 
as one confirms by splitting it into two parts: 
\[
\lim_{A\rightarrow\infty}\left\{\frac{e^{-A}}{2}\int_1^{B}e^tf(t)dt + 
\frac{e^{-A}}{2}\int_{B}^Ae^tf(t)dt\right\},
\]
where $B$ is a large positive constant. The first of the two
parts is evidently zero, for the second one notes that the
argument can be approximated as 
$e^t f(t) \approx -3e^t/(\theta^2t^2)$ for large $t >B$, whose 
absolute value has the upper bound $3e^A/(\theta^2A^2)$.
Thus in the limit the second part vanishes as well. 
Using (\ref{cc}) this means the constant $K$ in (\ref{h1sol}) 
can be computed as
\begin{equation}  
\label{K}  
K = \int_1^{\infty} (\frac{e^{-t}}{2}-1)f(t) dt
+ \int_0^1 (\cosh(t)-1)f(t) dt .
\end{equation}  

We are now in a position to determine the constant $F$ in 
(\ref{match}). Comparing (\ref{match}) to (\ref{exp}), we 
know that for small $x$ the first order contribution $h_1'$ 
must have a logarithmic singularity of the form 
\begin{equation}  
\label{hp}  
h_1'(x)\approx 3/\theta_{ap}^2\ln(x) + C(\theta/\theta_{ap})/\theta_{ap}^2.
\end{equation}  
Namely, for small $x$
\[
9Ca\ln(x) + F \approx  h_{out}'^3(x) \approx (h_0'(x)+Ca h_1'(x))^3 \approx
\theta_{ap}^3+9Ca\ln(x)+3Ca C(\theta/\theta_{ap})+O(Ca^2),
\]
so that $F=\theta_{ap}^3 + 3Ca C(\theta/\theta_{ap})$. 
Analysis of (\ref{h1sol}) for $x\rightarrow 0$ gives, using (\ref{K})
and $f(t)\approx -3/(\theta_{ap}t)^2$ for small $t$, 
\begin{equation}  
\label{C}  
C(\theta/\theta_{ap}) = \int_1^{\infty} (1-e^{-t})f(t)\theta_{ap}^2 dt + 
\int_0^1 \left[(1-e^{-t})f(t)\theta_{ap}^2+3/t\right] dt + 3.
\end{equation}  
This completely determines the the outer solution (\ref{exp}). 
Comparing it to the inner solution (\ref{voinov}) we finally obtain
\begin{equation}  
\label{ap}  
\theta_{ap}^3 + 3Ca C(\theta/\theta_{ap}) = \theta_e^3 - 9Ca\ln(L),
\end{equation}  
which is an equation to be solved for the apparent contact angle
$\theta_{ap}$. 

\begin{figure}
\includegraphics[width=0.6\hsize]{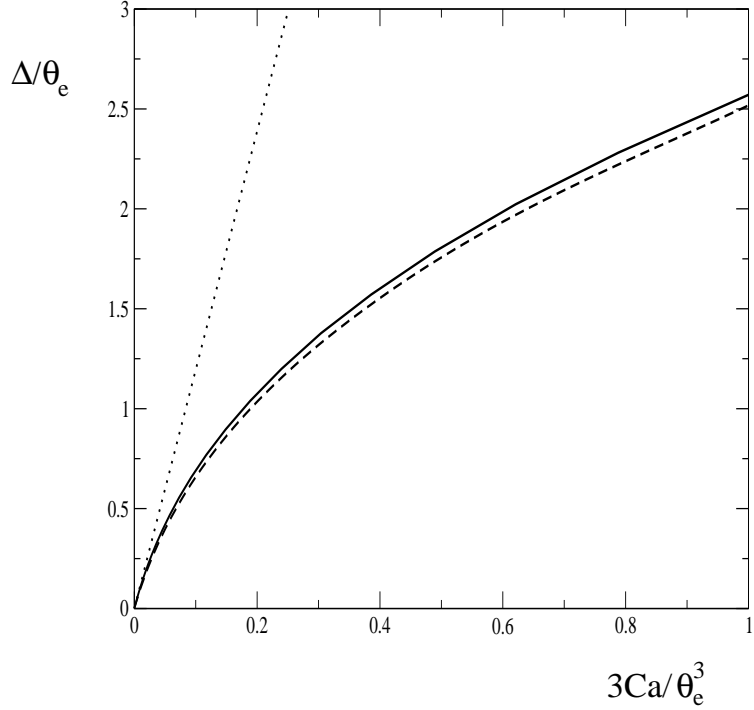}
\caption{\label{app_ad} 
The depression $\Delta$ of the meniscus (divided by the equilibrium contact 
angle $\theta_e$), as function of the reduced capillary number 
$3Ca/\theta_e^3$. The slip parameter is $3\lambda/\theta_e = 10^{-5}$,
$\theta/\theta_e = 1$. The full line is the result of a numerical
solution of (\ref{lub}), the dashed line is our theoretical 
result (\ref{ap}), using $\Delta = \theta_{ap}-\theta$. The dotted line
is the theoretical result of \cite{GHL90}.
   }
\end{figure}
\begin{figure}
\includegraphics[width=0.7\hsize]{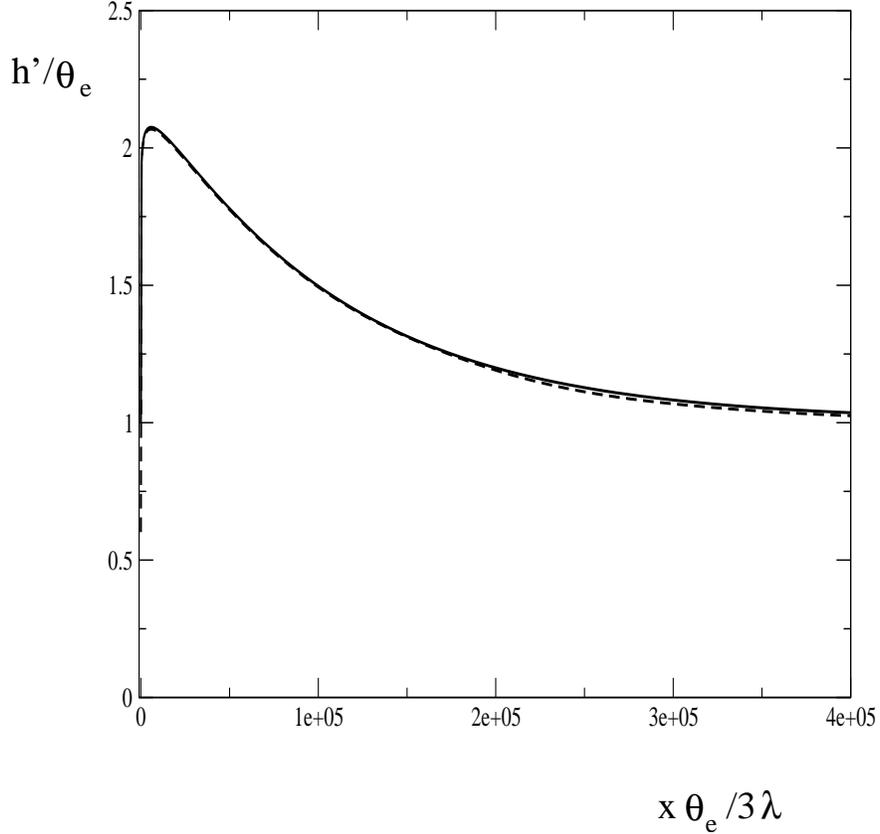}\caption{\label{compfiga} 
A comparison between a profile obtained numerically by integrating 
(\ref{lub}) at $\delta=3Ca/\theta_e^3=0.3$ and the outer 
solution (\ref{exp}). The other parameters are those of Fig.\ref{app_ad}.
For $\delta=0.3$, (\ref{ap}) gives $\theta_{ap}/\theta_e=2.32$.
   }
\end{figure}

To test the result of our matching procedure (\ref{ap}), we compare the 
depression $\Delta=\theta_{ap}-\theta$ of the meniscus with the 
result of a numerical solution of the original equation (\ref{lub}). 
Remarkably, the prediction, which contains no adjustable parameters,
remains extremely good up to a reduced capillary number of 
$3Ca/\theta_e^3 = 1$. To further appreciate the 
quality of the agreement, we plotted the result of an earlier 
theory \cite{GHL90} as the dotted line. In this earlier theory, a perturbation 
expansion of (\ref{lub}) in $Ca$ is matched directly to a linearized version
of (\ref{voinov}). However, the expansion is performed around the 
{\it static} profile corresponding to zero speed, whereas 
our expansion is around $h_0(x)$, which already 
incorporates a speed-dependent deformation of the surface. 
For the special case $\theta = \theta_e$, the result
of \cite{GHL90} is 
\begin{equation}  
\label{ghl}  
\Delta=-\theta_e\delta\left[\ln(L) - \gamma\right],
\end{equation}  
where $\gamma$ is Euler's constant. 
A comparison of the {\it profile} $h'(x)$ obtained numerically with
the outer solution (\ref{exp}) shows equally good agreement, as
shown in Fig.\ref{compfiga}. It is only for 
$x/\lambda \mbox{\ \raisebox{-.9ex}{$\stackrel{\textstyle <}{\sim}$}\ } 1$
that the outer solution begins to fail, since it has a logarithmic 
singularity at the origin.

\section{Pulling}

We now turn to the opposite case of a plate being withdrawn from a 
bath, for which the left-hand side of (\ref{lub}) is now positive.
The dominant balance close to the contact line is again
between the term on the left of (\ref{lub}) and the first 
term on the right. With the similarity transformation (\ref{scal}),
this is converted into 
\begin{equation}  
\label{simd}  
\frac{\delta}{H^2 + H} = H''',
\end{equation}   
which differs from (\ref{sim}) only by a sign. However, the behavior 
of solutions of (\ref{simd}) as $\xi\rightarrow\infty$ 
is completely different. 
This is best appreciated by considering the form of (\ref{simd})
valid away from the contact line where $H \gg 1$:
\begin{equation}  
\label{sima}  
\frac{1}{y^2} = y''',
\end{equation}   
where we have put $H(\xi) = \delta^{1/3}y(\xi)$. 

Remarkably, this equation has an exact solution, whose properties have 
been summarized in \cite{DW97}. In parametric form, a solution with
$y(0)=0$ reads
\begin{eqnarray}
\label{par}
\left.\begin{array}{l}
\xi = \frac{2^{1/3}\pi Ai(s)}{\beta(\alpha Ai(s) + \beta Bi(s))} \\    
y = \frac{1}{(\alpha Ai(s) + \beta Bi(s))^2}
                 \end{array}\right\}s\in [s_1\infty[,
\end{eqnarray}
where $Ai$ and $Bi$ are Airy functions \cite{AS}. 
The limit $\xi\rightarrow 0$ corresponds to $s\rightarrow \infty$, 
the opposite limit $\xi\rightarrow \infty$ to 
$s\rightarrow s_1$, where $s_1$ is a root
of the denominator of (\ref{par}):
\begin{equation}  
\label{s1}  
\alpha Ai(s_1)+\beta Bi(s_1)=0. 
\end{equation}  
Since the solution extends to
$s=\infty$, $s_1$ has to be the {\it largest} root of (\ref{s1}).

From (\ref{par}), the behavior of $y(\xi)$ for large $\xi$ can be
obtained \cite{DW97} (note that there is a misprint in equation (12)
of \cite{DW97}):
\begin{equation}  
\label{curv}  
y'(\xi) = \kappa_y \xi + b_y + O(\xi^{-1}), 
\end{equation}  
where 
\[
\kappa_y = \left(\frac{2^{1/6}\beta}{\pi Ai(s_1)}\right)^2, \quad
b_y = \frac{-2^{2/3}Ai'(s_1)}{Ai(s_1)}.
\]

The constant $\beta$ can be determined by matching (\ref{par}), which is
valid only for 
$\xi \mbox{\ \raisebox{-.9ex}{$\stackrel{\textstyle >}{\sim}$}\ } 1$,
to a solution of (\ref{simd}), which 
includes the effect of the cutoff and is thus valid down to the position
$\xi=0$ of the contact line. The limit of (\ref{par}) for small 
values of $\xi$ gives \cite{DW97}
\begin{equation}  
\label{small}  
H'^3(x) = \delta y'^3(\xi) \approx 
3\delta\ln(\pi/(2^{2/3}\beta^2\xi),
\end{equation}  
which remains valid for 
$\xi \mbox{\ \raisebox{-.9ex}{$\stackrel{\textstyle <}{\sim}$}\ } 
\beta^{-2}$. Thus, (\ref{small})
is a valid solution of the {\it full} equation (\ref{simd}) for
$1 \mbox{\ \raisebox{-.9ex}{$\stackrel{\textstyle <}{\sim}$}\ } 
\xi \mbox{\ \raisebox{-.9ex}{$\stackrel{\textstyle <}{\sim}$}\ } 
\beta^{-2}$. 

Following \cite{H83,E041}, we compare (\ref{small}) to the  
expansion of the full equation (\ref{simd}) in $\delta$. Using the 
boundary conditions $H(0)=0$ and $H'(0)=1$, one finds
\begin{equation}  
\label{sexp}  
H'(\xi)=1+\delta\left[\xi(\ln(\xi)-\ln(\xi+1)) - 
\ln(\xi+1)+C\xi\right] + O(\delta^2).
\end{equation}  
Since this solution has to match the logarithmic behavior (\ref{small}), 
we put the constant of integration $C$ to zero, and (\ref{sexp}) becomes 
for $\xi \gg 1$ 
\begin{equation}  
\label{lexp}  
H'^3(\xi)=1-3\delta\ln(\xi) + O(\delta^2).
\end{equation}  
Thus, comparing (\ref{lexp}) to (\ref{small}), we find
\begin{equation}  
\label{beta}  
\beta^2=\pi\exp(-1/(3\delta))/2^{2/3} + O(\delta). 
\end{equation}  
For $\delta\ll 1$, $\beta$ is indeed exponentially small, 
and (\ref{par}) has a logarithmic dependence over a large range 
of $\xi$-values:
$1 \mbox{\ \raisebox{-.9ex}{$\stackrel{\textstyle <}{\sim}$}\ } 
\xi \mbox{\ \raisebox{-.9ex}{$\stackrel{\textstyle <}{\sim}$}\ } 
\exp(1/(3\delta))$. This makes it possible to match (\ref{par}) to 
(\ref{lexp}) in the limit $\xi\gg 1$, although the ultimate 
behavior of (\ref{par}) for large $\xi$ is given by (\ref{curv}).

We are now in the position to match the {\it inner} solution 
of (\ref{simd}), which has the form (\ref{scal}), to an 
appropriate {\it outer} solution, following the prescription 
(\ref{match_cond}) given in the introduction. To achieve this 
matching it is enough to consider $h_{out}(x) = h_0(x)$ as 
given by (\ref{h0sol}), since (\ref{curv}) does not contain 
any logarithmic term. For $x/\lambda \gg 1$,
the solution of (\ref{simd}) is well described by (\ref{par}),
so we have 
\begin{subequations}
\label{inout:all}  
\begin{eqnarray}
h_{out}(x) = \theta x + (\theta - \theta_{ap})(e^{-x}-1) \label{inout:a}\\
h_{in}(x) = 3\lambda\delta^{1/3} y(x\theta_e/(3\lambda)), \label{inout:b}
\end{eqnarray}
\end{subequations}
where $y(\xi)$ is given by (\ref{par}),(\ref{beta}). The matching 
procedure must supply us with the parameter $s_1$ in (\ref{par}),
which is yet to be determined. 

The matching works in the limit $\lambda\rightarrow 0$, 
for which we require that the expansions 
\begin{eqnarray}
\label{comp}  
\left.\begin{array}{l}
h_{out}'(x) = \theta_{ap} + (\theta-\theta_{ap})x + O(x^2) \\
h_{in}'(x) = \theta_e\delta^{1/3}\left[\kappa_y x\theta_e/(3\lambda) + b_y
 \right] + O(\lambda/x)
                 \end{array}\right\}
\end{eqnarray}
agree, or
\begin{subequations}
\label{mequ:all}  
\begin{eqnarray}
\theta_{ap} = \theta_e\delta^{1/3}b_y ,\quad \label{mequ:a}  \\
\theta-\theta_{ap} = \theta_e^2\delta^{1/3}\kappa_y/(3\lambda). \label{mequ:b} 
\end{eqnarray}
\end{subequations}

Eliminating $\theta_{ap}$ between the two equations (\ref{mequ:all})
we finally have 
\begin{equation}  
\label{det}  
\frac{\theta}{(3 Ca)^{1/3}} +
\frac{2^{2/3}Ai'(s_1)}{Ai(s_1)} =
\frac{\theta_e\exp(-\theta_e^3/(9Ca))}{3\cdot2^{1/3}\pi Ai^2(s_1)\lambda},
\end{equation}  
which should be read as an equation for $s_1$. Once $s_1$ is 
known, all remaining parameters in $h_{in}$ and $h_{out}$ can
be found. For example, from (\ref{mequ:a})
one finds the apparent contact angle to be 
\begin{equation}  
\label{app}  
\theta_{ap} = \frac{-2^{2/3}(3Ca)^{1/3}Ai'(s_1)}
{Ai(s_1)},
\end{equation}  

\begin{figure}
\includegraphics[width=0.6\hsize]{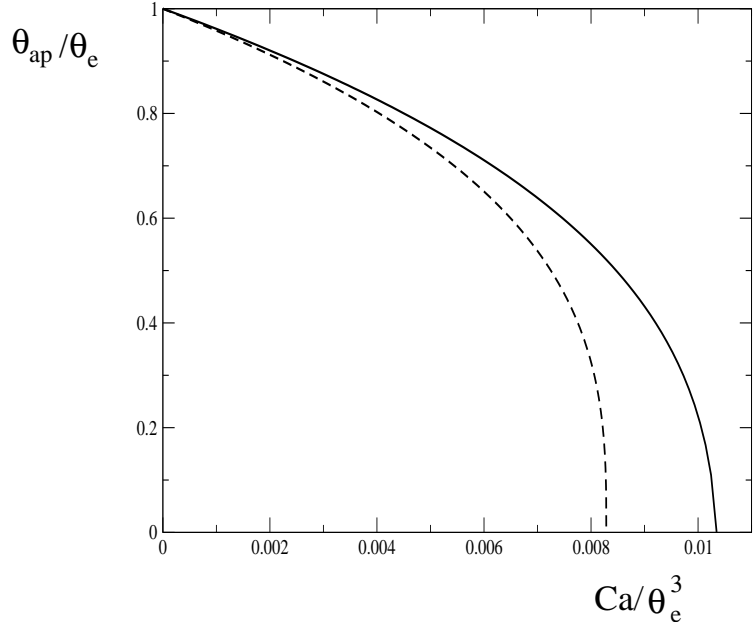}
\caption{\label{appfig} 
The apparent contact angle $\theta_{ap}$ according to (\ref{app}). 
The other parameters are $\lambda/\theta_e=10^{-6}$ and 
$\theta/\theta_e=2$. The dashed curve is the approximation 
(\ref{vdr}), valid for small capillary number. The apparent contact
angle goes to zero at the critical capillary number 
$Ca_{cr}/\theta_e^3=0.0103$.
   }
\end{figure}
an example of which is plotted in Fig. \ref{appfig}. At some finite
{\it critical} capillary number $Ca_{cr}$, $\theta_{ap}$ goes to zero. 
Physical solutions cannot exist for capillary numbers beyond 
that, since the outer solution only makes physical sense for 
$\theta_{ap} \ge 0$. Following \cite{E042},
we now derive a much simpler equation for the critical capillary 
number $Ca_{cr}$ at which the contact line disappears, and show 
that it is equivalent to $\theta_{ap}=0$. 

Since $\theta_{ap} \ge 0$, the {\it maximum} value of the left
hand side of (\ref{mequ:b}), which is the curvature of the outer
solution, is $\theta$. The right hand side, corresponding to the curvature
of the inner solution, is on the other hand bounded from 
below. From (\ref{curv})
it is seen that the minimum curvature corresponds to the condition that
$Ai(s_1)$ must be maximal among solutions of (\ref{s1}).
By choosing $\alpha=\alpha_{cr}\equiv-\beta Bi(s_{max})/Ai(s_{max})$
we can in fact ensure that $Ai$ assumes its global maximum 
$0.53566\dots$, which occurs for $s=s_{max}=-1.0188\dots$. 
Thus we have singled out a unique solution of (\ref{sima})
which minimizes the curvature 
\begin{equation}  
\label{ccr}  
\kappa^{cr}_{y} = \frac{\exp[-\theta_e^3/(9Ca)]}{2^{1/3}\pi (Ai(s_{max}))^2},
\end{equation}  
the value of which increases with capillary number as expected.

Now by equating the maximum value of the left hand side of 
(\ref{mequ:b}) with the minimum value of the right of (\ref{mequ:b}) 
we obtain an equation 
for a capillary number above which no solution can exist:
$\theta = \theta_e^2\delta^{1/3}\kappa^{cr}_y/(3\lambda)$,
or explicitly
\begin{equation}  
\label{dcr}  
Ca_{cr} = \frac{\theta_e^3}{9}\left[\ln\left(
\frac{Ca_{cr}^{1/3}\theta_e}{18^{1/3}\pi (Ai(s_{max}))^2\lambda\theta}
\right)\right]^{-1}.
\end{equation}  
But at the capillary number given by (\ref{dcr}), the first matching
condition (\ref{mequ:a}) is also satisfied identically, since 
$\theta_{ap} = b_y = 0$. This is because $Ai$ is extremal, so 
$Ai'=0$. Thus $Ca_{cr}$ as given by (\ref{dcr}) gives 
exactly the critical capillary number corresponding to $\theta_{ap} = 0$ in 
Fig.\ref{appfig}. This confirms a classical conjecture by 
Derjaguin and Levi \cite{DL64}, later reiterated by others 
\cite{BR79}, that the transition to a film is characterized by
the apparent contact angle going to zero. This criterion was 
confirmed experimentally in \cite{SP91}, using fibers being pulled
out of a viscous liquid. 

\begin{figure}
\includegraphics[width=0.6\hsize]{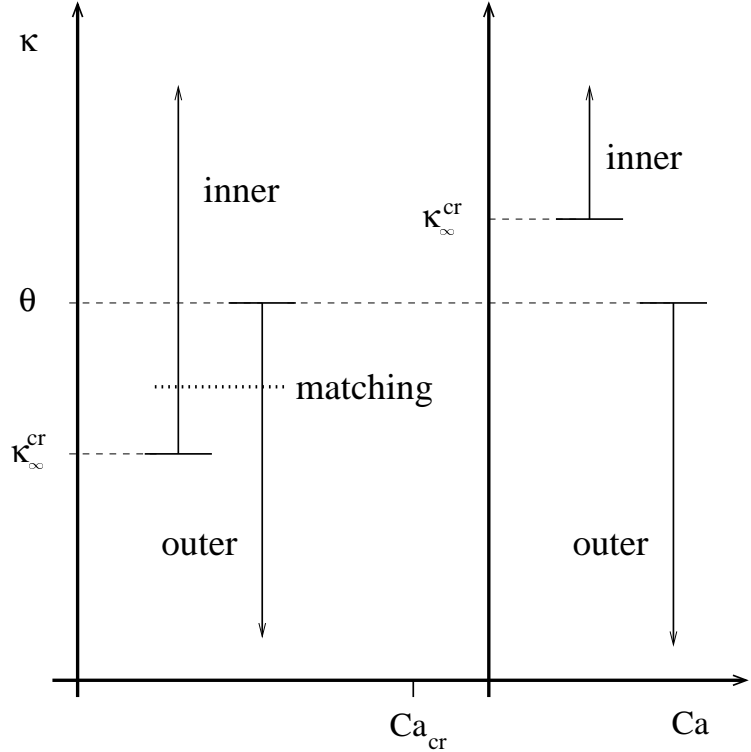}
\caption{\label{theory} 
A schematic illustrating the transition to a LLD film for a 
receding contact line. Below $Ca_{cr}$, the curvatures of 
all possible inner and outer solutions have an overlap region, 
so they can be matched. Which solution from the overlap region
is selected (illustrated by the dotted line), is determined by
(\ref{det}). Above $Ca_{cr}$ no matching is possible, since the 
the maximum curvature of all possible outer solutions lies below
the minimum curvature of all the inner solutions. 
   }
\end{figure}

The concept underlying equation (\ref{dcr}) for the critical 
capillary number is illustrated again in Fig.\ref{theory}.
Below the critical capillary number, there is some overlap 
in the curvature of the inner and the outer solutions (left
diagram in Fig.\ref{theory}). The matching equation (\ref{det})
determines which solution is selected from this overlap region.
Above the critical capillary number, 
there can be no solution (right diagram in Fig.\ref{theory}): 
there is no outer solution which could be matched 
to an inner solution, because this would result in a sudden jump in the 
curvature. 

\begin{figure}
\includegraphics[width=0.6\hsize]{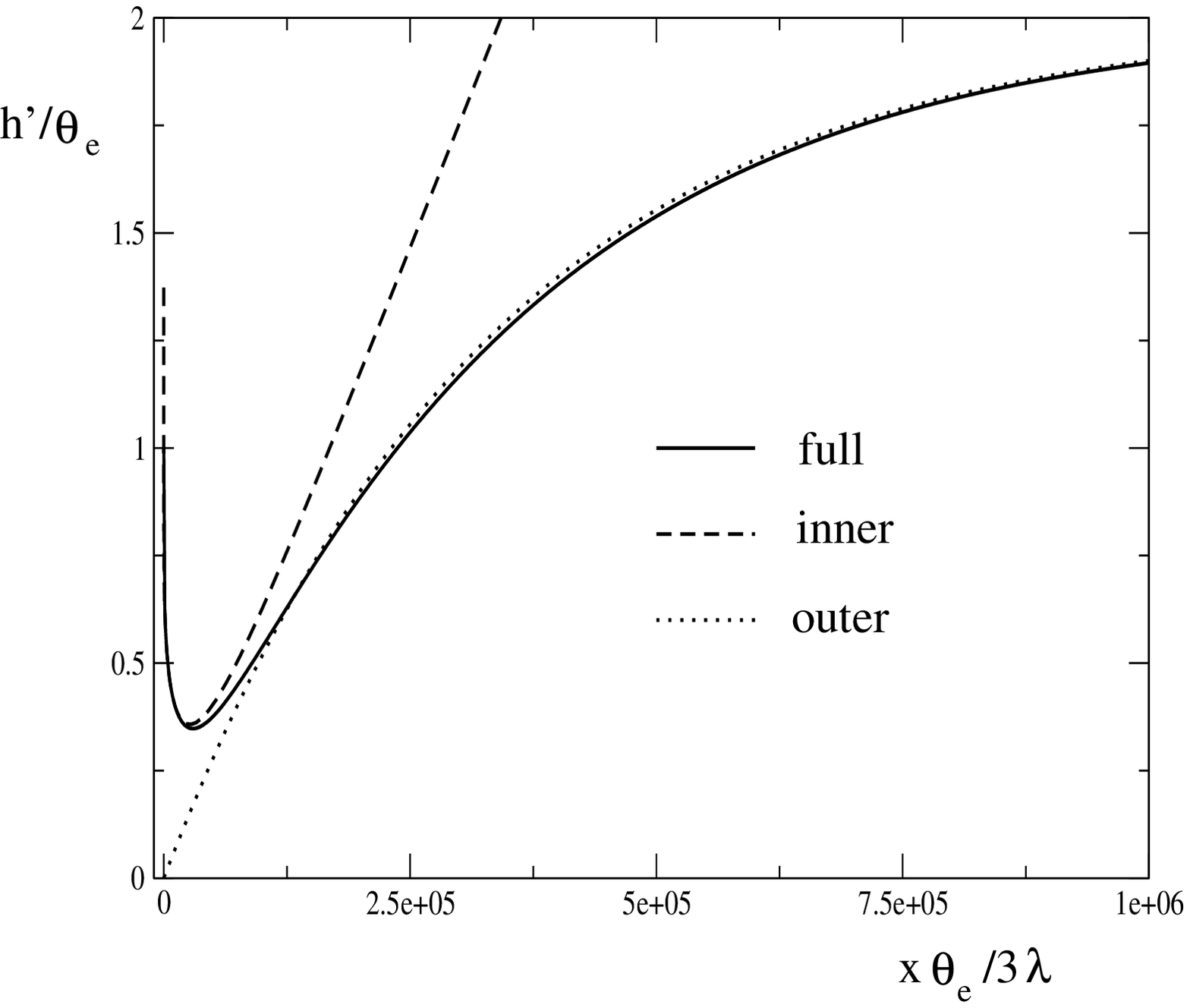}
\caption{\label{compare} 
A comparison of the full solution at the critical capillary number with
the inner and outer solutions. We plot the slope of the interface,
so $h'(0)/\theta_e=1$ for the full solution, and $h'(0)=0$ for the 
outer solution, consistent with the condition by Derjaguin and Levi. 
The other parameters are $\lambda/\theta_e=10^{-6}$ 
and $\theta/\theta_e=2$, so from (\ref{dcr}) we obtain 
$Ca_{cr}/\theta_e^3=0.01032$.
   }
\end{figure}

In Fig. \ref{compare} we show the result of a numerical integration
of (\ref{lub}) at the critical capillary, and compare it to the 
inner and outer solutions (\ref{inout:all}). The critical capillary 
number was found numerically by raising $Ca$ until no more solutions
of (\ref{lub}) could be found. For large $\xi$, the slope of the 
outer solution (\ref{inout:a}) agrees with the numerical solution, 
but extrapolates to $h'=0$ at the contact line, as reqired by 
$\theta_{ap}=0$. Coming from the interior, the inner solution 
(\ref{inout:b}) agrees with the full solution up to the turning 
point. 

\begin{figure}
\includegraphics[width=0.7\hsize]{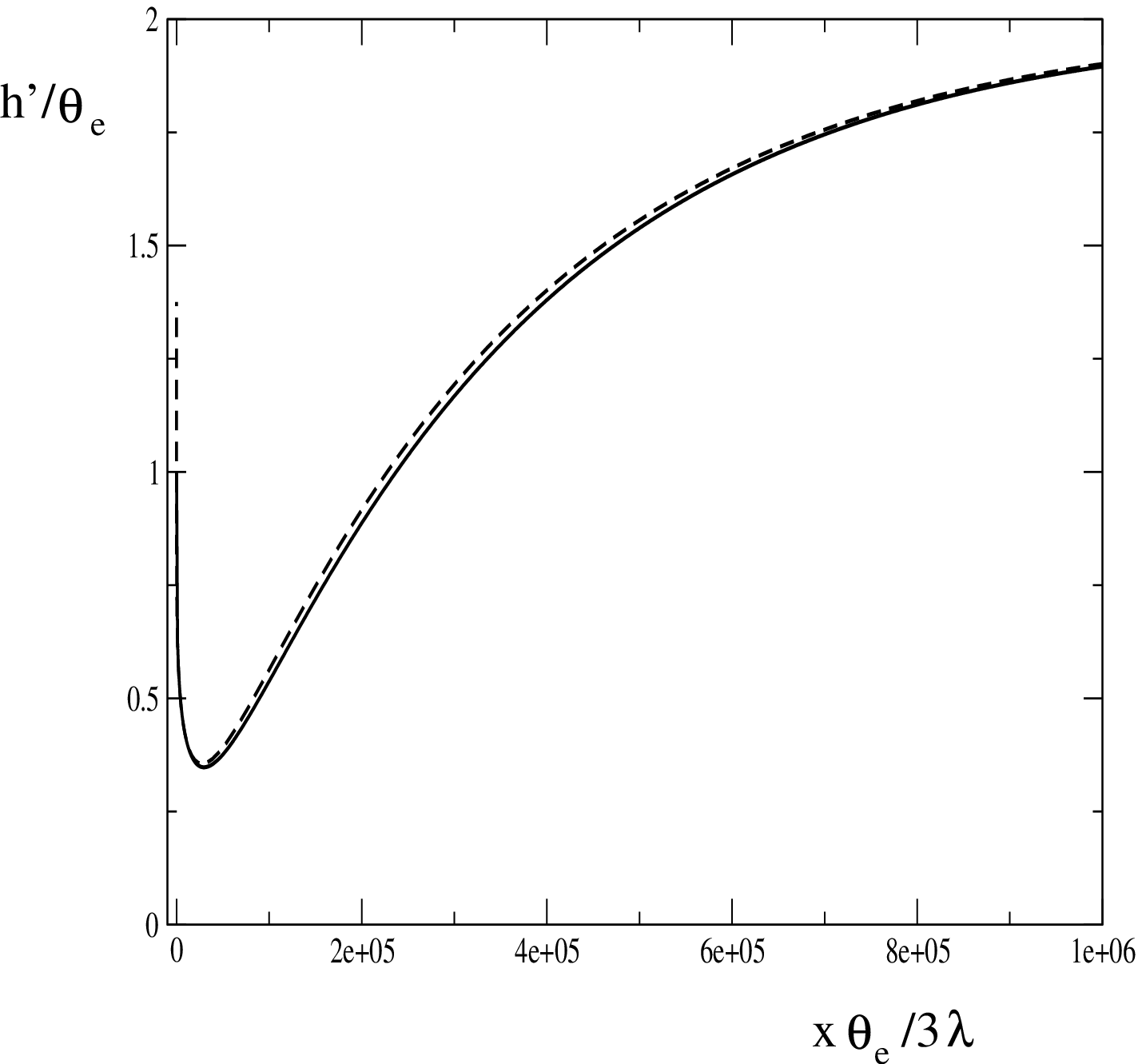}\caption{\label{compfig} 
The interface slope at the critical capillary number for 
$\theta/\theta_e=2$ and $\lambda/\theta_e = \cdot 10^{-6}$.
The full line is the result of the numerical integration at 
$Ca_{cr}/\theta_e^3=0.0103$, 
while the theoretical prediction (\ref{dcr}) gives
$Ca_{cr}/\theta_e^3=0.01032$. 
The dahes line is the composite solution (\ref{composite}).
   }
\end{figure}

To obtain a profile that is more uniformly valid, one can 
use a composite approximation \cite{H}. The idea is to 
add the inner and outer solutions that have been matched, and to subtract 
the behavior (\ref{comp}) in the region in which the two
solutions overlap:
\begin{equation}  
\label{composite}  
h_{comp}(x) = h_{in}(x) + h_{out}(x) - 
\left[\theta_{ap} + (\theta-\theta_{ap})x\right].
\end{equation}  
It is evident that $h_{comp}$ agrees with the full solution for
large {\it as well as} for small $x$, hence it will be the 
best global approximation at this order of the matching. This excludes 
the region 
$\xi \mbox{\ \raisebox{-.9ex}{$\stackrel{\textstyle <}{\sim}$}\ } 1$
where (\ref{inout:b}) has a logarithmic singularity. 
Figure \ref{compfig} illustrates the remarkable 
agreement of the composite solution with the full numerical result
at the critical capillary number, where the approximation is expected
to be {\it worst}. 

Having studied the critical capillary number, let us return 
once more to the matching condition (\ref{det}) which determines
the solutions for $Ca<Ca_{cr}$. Namely, it is instructive to 
obtain explicit solutions of (\ref{det}) in the limit of {\it small}
$Ca$, in which case its solution $s_1$ is large,
and the asymptotics of the Airy function \cite{AS} gives 
\begin{equation}  
\label{Airy}  
Ai(s) \approx e^{-2s^{3/2}/3}s^{-1/4}/(2\pi^{1/2}).
\end{equation}  
Thus, (\ref{det}) becomes for small $Ca$:
\begin{equation}  
\label{dets}  
\frac{\theta}{(3Ca)^{1/3}} - 2^{2/3}s^{1/2} - 
\frac{2^{5/3}\theta_e s^{1/2}}{3\lambda}
e^{-\theta_e^3/(9Ca)+4s^{3/2}/3} \approx 0,
\end{equation}  
and (\ref{app}) is
\begin{equation}  
\label{bsm}  
\theta_{ap}\approx (3Ca)^{1/3}2^{2/3}s^{1/2}. 
\end{equation}  
In the simplest approximation the exponent in (\ref{dets}) must
vanish, $s = \left(\theta_e^3/(12Ca)\right)^{2/3}$, and inserting
this into (\ref{bsm}) gives $\theta_{ap}\approx \theta_e$. 
Not surprisingly, at very small capillary numbers the interface is 
hardly deformed at all. 

To go beyond this approxiamtion, we
put $s^{1/2}=\left(\theta_e^3/(12Ca) + \Delta s\right)^{1/3}$, 
insert into (\ref{dets}), and analyze the result for small $Ca$,
which gives 
\[
\Delta s\approx \frac{3}{4}\ln
\left[\frac{3\lambda}{2\theta_e^2}(\theta-\theta_e)\right].
\]
Putting this into (\ref{bsm}) leads to the next approximation
\begin{equation}  
\label{vdr}  
\theta_{ap}^3\approx\theta_e^3-9Ca\ln
\left[\frac{2\theta_e^2(\theta-\theta_e)}{3\lambda}\right],
\end{equation}  
which is similar to equation (\ref{ap}) for the apparent contact
angle in the advancing case. Indeed, for small capillary number
(\ref{ap}) becomes 
\begin{equation}  
\label{apsmall}  
\theta_{ap}^3\approx\theta_e^3+9Ca\ln
\left[\frac{e\theta_e\exp(C(\theta/\theta_e)/3)}{3\lambda}\right],
\end{equation}  
which has the same form as (\ref{vdr}).
However, even for small capillary number the receding and the advancing
case needs to be treated differently, as the constant inside the 
logarithm is {\it different}. For $\theta=\theta_e$ the argument of the
logarithm in (\ref{vdr}) becomes singular, which means one has to 
go to an even higher 
order in the approximation, but we are not pursuing this special
case here. The approximation (\ref{vdr}) is plotted as the dashed
line in Fig.\ref{appfig}. If extrapolated naively to $\theta_{ap}=0$,
it gives a surprisingly good estimate of $Ca_{cr}$.

\section{Discussion}
Let us begin by considering rather straightforward extensions and 
generalizations of the present theory. Firstly, our arguments 
are not limited to a specific contact line model, since they are
based entirely on hydrodynamic arguments {\it away} from the contact 
line. For example, if van-der-Waals forces are dominant near the 
contact line \cite{GHL90}, this only changes the parameter $L$ 
appearing in (\ref{voinov}) (advancing contact angle) or $\beta$ in 
(\ref{small}) (receding contact line). In the latter case, the slip 
length $\lambda$ in (\ref{dcr}) has to be replaced by 
$\sqrt{A/(6\pi\gamma)}/(6\theta_e)$, where $A$ is the Hamaker constant. 
Our research also suggests that the calculation of the critical 
capillary number is captured fully by lubrication theory, even if 
$\theta_e$ is {\it not} small. In that case the flow directly at the 
contact line would have to be described without resorting to lubrication
theory \cite{V76,C86}, but the relevant region where matching occurs is 
characterized by slopes $h'(x)$ which {\it are} small.

Secondly, one can generalize to a different geometry. To this end
one has to replace (\ref{h0sol}) by the appropriate static solution 
for the problem at hand. This is done almost trivially for the case 
of a {\it vertical} plate or a fiber \cite{LL84}, in which case the
lubrication description (\ref{lub}) is no longer valid far from 
the contact line, since $\theta$ is not small. This however does not 
pose a problem since this part of the profile is determined by 
surface tension and gravity alone. In the same spirit, the 
present model can be exteded to a flow inside a capillary tube,
with only very minor changes to the value of the critical capillary
number \cite{H01}. 

Thirdly, one can consider dynamical effects, of particular
interest for the unstable case of a receding  contact line. 
In \cite{H01} it was found that when the stationary profile
vanishes, it is not followed directly by the LLD film, which is 
of macroscopic thickness. Rather, there is a narrow range of 
speeds where the contact line is pulled up the plate, but
at a speed that is {\it smaller} than $U$, i.e. the contact line
is partially slipping. The thickness of the film that is left 
behind is in the order of $\lambda$, i.e. {\it microscopic}. 
Only when the speed is raised still further does the LLD film 
appear. So far these results are only numerical; a full analytical 
theory would be desirable. Note also that these transitions
are strongly hysteretical. Once the LLD film has appeared,
it can be sustained to much lower capillary numbers than $Ca_{cr}$. 
It is usually assumed that the LLD film vanishes when its thickness
has reached the range of intermolecular forces \cite{Q99}, but
we are not aware of any theoretical investigation of this problem.

Next we come to the experimental evidence. The most extensive 
experiments on the critical capillary number for a
receding contact line were done with a capillary tube \cite{Q91},
by pushing out a viscous liquid. Using a variety of different 
materials it was found that instability occurs at a given 
value of the reduced capillary number $Ca/\theta_e^3$, in agreement
with the present theory. The actual value of the critical capillary 
number, however, is about a factor of two too low, if the
theoretical estimates are based on $\lambda\sim$ nm. Several
possible explanations suggest themselves. First, the materials
used in \cite{Q91} have considerable contact angle hysteresis, 
pointing to surface roughness. This will tend to reduce the critical 
capillary number \cite{GR01,GR03}. In addition, any speed dependence
of the microscopic contact angle, neglected in the present description,
will effectively lower $\theta_e$ and thus lead to a smaller
critical capillary number. 

Our theory for the vanishing of the receding contact line has
some similarities with an earlier theory \cite{G86},
in that the critical capillary number is proportional to 
$\theta_e^3$ (cf. (\ref{dcr})). However, it differs in 
predicting a vanishing apparent contact angle at the transition, 
while it is $\theta_{ap}/\theta_e=1/\sqrt{3}$ in \cite{G86}.
The approach of \cite{G86} is also different in that it considers
the local problem in isolation, hence the dependence of 
$Ca_{cr}$ on parameters of the outer problem like $\theta$
cannot be captured. In fact, we believe that the mechanism 
for instability proposed in \cite{G86}, which is based on 
an approximate solution of (\ref{simd}), contains a flaw. 
In \cite{E041} we use the case of the advancing contact 
angle (\ref{sim}) to show that the method of solution 
proposed in \cite{G86} cannot correctly predict the nonlinear dependence 
of the angle on speed. But it is precisely this nonlinear 
dependence which lies at the heart of the stability analysis of 
\cite{G86}. 

The most important extension of the present theory of 
contact line instability however is its application to 
higher dimensions, in which the contact line no longer 
remains straight. If the plate withdrawn from the liquid 
bath is sufficiently wide, the contact line inclines relative
to its direction of motion. Two sections of the contact line
that have inclined in opposite senses meet at a sharp corner, 
so that the whole contact line is serrated in an irregular 
fashion \cite{BR79}. 

A more controlled recent experiment is that 
of a viscous drop running down an inclined plane \cite{PFL01}.
At a critical speed, the initially rounded tail of the drop forms
a sharp corner. 
A recent theory \cite{LS04} explains the drop profile near
the corner of the drop, but not the critical speed at which 
the the corner first occurs, nor its opening angle. To give a 
complete description of the transition, the microscopic 
neighborhood of the contact line has to be included, as was done 
in the present theory for a straight contact line. One important 
difference between the case of a sliding drop and that of
the present paper is that recent experimental evidence suggests
that the transition occurs at a {\it finite} value of the apparent 
contact angle \cite{L04}. Studies to understand the three-dimensional
instability of a sliding drop are currently under way. 

\acknowledgments
I am grateful to Martin Sieber for several important discussions
on the matching procedure, to Seth Lichter for improving my
physical understanding, and to Lorena Barba for a careful reading 
of the manuscript. 

\end{document}